\documentclass[12pt]{article}
\usepackage{epsfig}
\usepackage{graphics}
\usepackage{psfrag}
\usepackage{amsmath}
\usepackage{amssymb}
\usepackage{color}

\newcommand{\bea}{\begin{eqnarray}}
\newcommand{\eea}{\end{eqnarray}}
\newcommand{\simgt}{\hbox{ \raise3pt\hbox to 0pt{$>$}\raise-3pt\hbox{$\sim$} }}
\newcommand{\simlt}{\hbox{ \raise3pt\hbox to 0pt{$<$}\raise-3pt\hbox{$\sim$} }}

\newcommand{\clfn}{\setcounter{footnote}{0}}

\newcommand{\LQ}{\Lambda_{\rm QCD}}

\begin{document}

\begin{titlepage}

    \begin{flushright}
      \normalsize UT--15--20\\
      \normalsize TU--997\\
      \today
    \end{flushright}

\vskip2.5cm
\begin{center}
\Large\bf\boldmath
Strong IR Cancellation in Heavy Quarkonium\\ 
and Precise Top Mass Determination
\unboldmath
\end{center}

\vspace*{0.8cm}
\begin{center}
{\sc Y. Kiyo}$^{a}$,
{\sc G. Mishima}$^{b}$ and
{\sc Y. Sumino$^{c}$}\\[5mm]
  {\small\it $^a$ Department of Physics, Juntendo University}\\[0.1cm]
  {\small\it Inzai, Chiba 270-1695, Japan}

  {\small\it $^b$ Department of Physics, University of Tokyo}\\[0.1cm]
  {\small\it  Bunkyo-ku, Tokyo 113-0033, Japan}

  {\small\it $^c$ Department of Physics, Tohoku University}\\[0.1cm]
  {\small\it Sendai, 980-8578 Japan}

\end{center}

\vspace*{2.8cm}
\begin{abstract}
\noindent
Combining recent perturbative analyses on the static QCD potential
and the quark pole mass, we find that, for the
heavy quarkonium states $c\bar{c}$, $b\bar{b}$ and $t\bar{t}$,
(1) ultra-soft (US) corrections in the binding energies are small, and
(2) there is a stronger cancellation of IR contributions
than what has been predicted by 
renormalon dominance hypothesis.
By contrast, for a hypothetical heavy quarkonium
system with a small number of active quark flavors
($n_l\approx 0$),
we observe evidence that renormalon
dominance holds accurately 
and that non-negligible contributions from
US corrections exist.
In addition, we examine contributions of renormalons at $u=- 1$.
As an important consequence, 
we improve on a previous prediction for
possible achievable accuracy of top quark 
$\overline{\rm MS}$--mass measurement
at a future linear collider and estimate that in principle
20--30~MeV accuracy is reachable.
\vspace*{0.8cm}
\noindent

\end{abstract}


\vfil
\end{titlepage}

\newpage

During the past few decades, there have been significant developments
in the analysis of heavy quarkonium systems using perturbative
QCD.
Developments in computational technology greatly advanced
our understanding on the nature of quark masses and interquark
forces.
We anticipate that eventually these developments will
deepen our understanding on the structure of
perturbative QCD in more general contexts.


Recently an important step toward this direction has been achieved.
A computation 
was completed of the four-loop relation between
the quark pole mass and the 
mass in the modified-minimal-subtraction scheme
($\overline{\rm MS}$ mass) \cite{Marquard:2015qpa}.
This result, when combined with other known results
such as the three-loop correction ($a_3$) to the static
QCD potential $V_{\rm QCD}(r)$ \cite{Anzai:2009tm},
sets our analysis at a new stage,
namely, at full next-to-next-to-next-to-leading order
(NNNLO) in terms of short-distance
quark masses.
It realizes a cancellation of infra-red (IR) dynamics at this order.

In this first analysis we report
what can be learned by combining existing results.
In particular we compare the results of 
\cite{Sumino:2013qqa,Ayala:2014yxa,Marquard:2015qpa}
to make clearer the nature of the perturbative series
of the heavy quarkonium energies,
concerning (1) corrections from the ultra-soft (US)
energy scale and (2) the renormalon dominance hypothesis.
In addition, we examine contributions of an ultra-violet
(UV) renormalon 
at $u=- 1$ and discuss possible contributions of an IR renormalon
at $u=+1$.

Motivations for performing such an analysis can be
stated as follows.
A few years ago, a convincing evidence has been presented for
the existence of IR renormalons in the perturbative series
of the energy of a static color
source, which has an IR structure common to the quark pole
mass \cite{Bauer:2011ws}.
Hence, it is among general interests how accurately the
renormalon dominance picture holds for the quark pole mass.
Furthermore, contributions of US corrections to the 
quarkonium energy have collected attention since long
time \cite{Appelquist:es,Voloshin:1978hc,Brambilla:1999xf}.
Despite an original expectation of being dominating
at IR,
there have been evidences that US corrections are
moderate in size
from comparisons of
the perturbative predictions with experimental data for
the bottomonium spectrum \cite{Brambilla:2001fw,Brambilla:2001qk}, 
phenomenological potential models
of heavy quarkonia \cite{Sumino:2001eh}, and
lattice computations
of $V_{\rm QCD}(r)$ \cite{Necco:2001xg}.
However, extraction of an accurate size of the US corrections still
remains a challenge \cite{Bazavov:2014soa}.

Important applications of this type of analysis include precise
determination of the masses of the heavy quarks $c$, $b$ and $t$
from the energy levels of the lowest-lying heavy quarkonium
states \cite{RecentBottomMassDetermination}. 
(For earlier works, 
see \cite{Brambilla:2004wf} and references therein.)
In this paper we apply our new understanding to a study of
the possible achievable accuracy of 
top quark mass measurement expected
at a future linear collider.
Today, a precise determination of the top quark mass
is highly demanded, for a precision test of the standard model 
of particle physics (SM) \cite{Baak:2012kk},
and also since
the top quark mass plays a crucial role in
the vacuum stability of the SM
at a very high energy scale \cite{Degrassi:2012ry}.
Hence, progress in our understanding of 
the heavy quarkonium
states may lead to an access to deep aspects of the
SM.

The pole-$\overline{\rm MS}$
mass relation can be expressed in a series expansion in
the strong coupling constant as 
\begin{eqnarray}
&& 
m_{\rm pole}
=
\overline{m}\,
\left[ 1 + d_0 \frac{ \alpha_s(\overline{m})}{\pi}
         +  d_1 \left(\frac{ \alpha_s(\overline{m})}{\pi}\right)^2
         +  d_2  \left(\frac{ \alpha_s(\overline{m})}{\pi}\right)^3
         +  d_3  \left(\frac{ \alpha_s(\overline{m})}{\pi}\right)^4
         + {\cal O} (\alpha_s^5 )
\right] \, .
\nonumber\\
\label{m-pole}
\end{eqnarray}
Here,
$\overline{m} \equiv m_{\overline{\rm MS}}(m_{\overline{\rm MS}})$
denotes the $\overline{\rm MS}$ mass renormalized at the
$\overline{\rm MS}$ mass scale;
$\alpha_{s}(\mu)=\alpha_s^{(n_l)}(\mu)$ represents
the strong coupling constant in the $\overline{\rm MS}$ scheme, 
where $n_l$ is the number of massless quark flavors
($n_l=3$, 4 and 5 for the charm, bottom and top quarks, respectively);
the renormalization scale $\mu$ is set to $\overline{m}$.
In most part of this paper,
we use the
coupling constant of the theory with
$n_l$ flavors only as the expansion parameter.
The coefficients $d_i$ can be 
obtained from the corresponding mass 
relations in the full theory (with $n_h$ heavy quarks and $n_l$ light quarks),
respectively, by rewriting them in terms of the coupling constant of the theory with
$n_l$ light quarks only.

Let us first summarize the results of the previous analyses,
on which our analysis is based.
Refs.~\cite{Sumino:2013qqa,Ayala:2014yxa} estimated $d_3$ on the basis of different
assumptions, prior to Ref.~\cite{Marquard:2015qpa}, which 
accomplished the exact
computation of $d_3$:
\begin{itemize}
\item
Ref.~\cite{Sumino:2013qqa} required stability of the perturbative
prediction for $2m_{\rm pole}+V_{\rm QCD}(r)$
at relatively large $r$.
Essentially the only assumption made is that US corrections 
in $V_{\rm QCD}(r)$ do not deteriorate perturbative
stability (which holds up to NNLO) at NNNLO.
\item
Ref.~\cite{Ayala:2014yxa} assumed renormalon dominance in 
$m_{\rm pole}$ and $V_{\rm QCD}(r)$ and estimated their
contributions from the latter.
Contribution of US corrections in $V_{\rm QCD}(r)$
was subtracted
in this estimate.\footnote{
Since
US corrections in $V_{\rm QCD}(r)$ do not
contribute to the renormalon at $u=1/2$, this
manipulation is justified within the renormalon 
dominance hypothesis.
}
\item
The exact values of $d_3$ are obtained combining
the results of direct
perturbative computations in
\cite{Beneke:1994qe,Lee:2013sx,Marquard:2015qpa}.
\end{itemize}
\begin{table}
\begin{center}
\begin{tabular}{c|lllllll}
\hline
~~$n_l$ & ~~~0 & ~~~1 & ~~~2 & ~~~3 & ~~~4 & ~~~5 & ~~~6
\\
\hline
\small $d_3^{n_h=1}$ & 
\small 
$3556.5$ & \small $2853.4$ & \small $2232.9$ & \small $1691.2$ & \small $1224.0$ & \small $~827.4$ & \small $~497.2$
\\
\hline
\end{tabular}
\caption{\small
Exact result of $d_3$ for $0\leq n_l \leq 6$ in the full theory,
with $n_h=1$ heavy quark and $n_l$ massless quarks.
We use 
eq.~(\ref{d3exact-full-theory})
obtained by a fit
of the results in \cite{Beneke:1994qe,Lee:2013sx,Marquard:2015qpa}.
An error of $\pm 21.5$ is assigned to each value.
}
\label{d3FullTheory}
\end{center}
\end{table} 

Only the values for $n_l=3,4,5$ are presented explicitly
in the final form
in \cite{Marquard:2015qpa} (for the full theory with $n_h=1$).
Since we need the values for other $n_l$'s in our analysis,
we derive the
exact result of $d_3$ given as a cubic polynomial of
$n_l$ as
\bea
d^\text{exact}_{3,\text{ full theory}}= 
-0.67814 \, n_l^3 + 43.396 \, n_l^2- 745.85 \, n_l
+ 3556.5 \, ,
\label{d3exact-full-theory}
\eea
where an error of $\pm 21.5$ is assigned to its value for each $n_l$.
We determined
the last two coefficients of eq.~(\ref{d3exact-full-theory}) by a fit
using the results of \cite{Beneke:1994qe,Lee:2013sx}
in addition to the result of \cite{Marquard:2015qpa}.
For the reader's convenience, we list
the exact result of $d_3$ in the full theory
in Tab.~\ref{d3FullTheory} using this formula
for $0\leq n_l \leq 6$.

As already mentioned, we convert
the above formula using the coupling of
the theory with $n_l$ massless quarks only
as the expansion parameter.
This gives
\bea
d_{3,\text{ converted}}^\text{exact}= -0.67814 \, n_l^3 + 43.396 \, n_l^2-745.42 \, n_l
+3551.1 \, ,
\label{d3exact}
\eea
with the same error $\pm 21.5$.
In the rest of the analysis, we use this 
$d_3$ for various $n_l$'s.

\begin{figure}[t]
\begin{center}
\includegraphics[width=14cm]{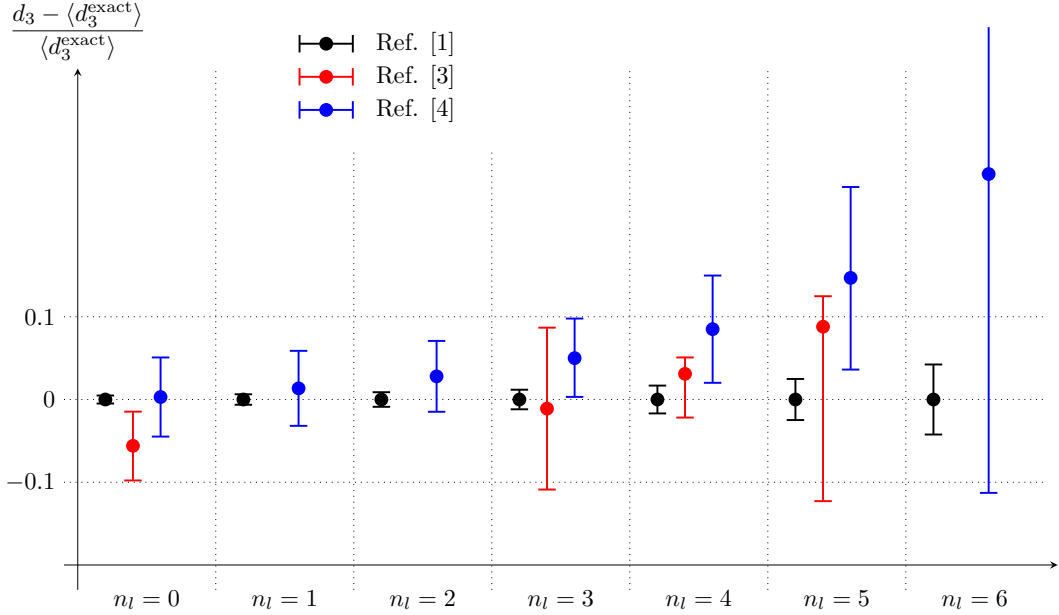}
\end{center}
\vspace*{-7mm}
\caption{\small
Comparison of $(d_3 - \langle d_3^\text{exact}\rangle)/\langle d_3^\text{exact}\rangle$ for
the (converted) exact value of $d_3$ 
and the two estimates, where
$\langle d_3^\text{exact}\rangle$ denotes the central value
of $d_3^\text{exact}$.
\label{ErrorBars}}
\end{figure}

In Tab.~\ref{results} we summarize the two estimates and the
exact result for $0\leq n_l\leq 6$.\footnote{
Since we use the converted $d_3^\text{exact}$, its values for $n_l=3,4,5$
listed in this table are different from TABLE III of 
\cite{Marquard:2015qpa}.
In this sense, the comparison in TABLE III of \cite{Marquard:2015qpa}
is not consistent, since $d_3$'s in the different definitions
are compared.
Numerically 
the differences due to different definitions
are small, nonetheless.
}
The relative accuracies
are compared visually in Fig.~\ref{ErrorBars}.
Overall, we find a reasonable agreement of the previous estimates
and the exact results, with respect to the assigned errors.
The relative accuracies of the estimates are also fairly good,
at order 10\% level.
These features provide certain justification to the
used assumptions in these estimates.

\begin{table}[t]
\hspace{-15mm}
\begin{tabular}{l|lllllll}
\hline
~~$n_l$ & ~~~~0 & ~~~~1 & ~~~~2 & ~~~~3 & ~~~~4 & ~~~~5 & ~~~~6
\\
\hline
\small $d_3^\text{est}$\cite{Sumino:2013qqa} & \small $3351(152)$
& \small ~~~~--- & \small ~~~~--- & \small 1668(167) & \small $1258^{+26}_{-66}$& \small
$897^{~+31}_{-175}$& \small ~~~~---
\\
\small $d_3^\text{est}$\cite{Ayala:2014yxa} & \small
$3562(173)$ & \small 2887(133) & \small 2291(98) & \small 1772(82) & \small 1324(81) & \small 945(92) & \small 629(191)
\\
\small $d_3^\text{exact}$\cite{Marquard:2015qpa} & \small 
$3551.1(21.5)$ & \small 2848.4(21.5) & \small 2228.4(21.5) & \small 1687.1(21.5) & \small 1220.3(21.5) & \small 824.1(21.5) & \small 494.3(21.5)
\\
\hline
\end{tabular}
\caption{\small
Summary table of relevant estimates and exact results 
of $d_3$.
The first line shows the estimates based on stability of
the perturbative
prediction for $2m_{\rm pole}+V_{\rm QCD}(r)$;
the second line shows the estimates based on renormalon dominance
hypothesis;
the third line shows the exact results (converted to the values in the
$n_l$ flavor theory).
}
\label{results}
\end{table} 

Furthermore, we can make a closer examination.
In particular, 
the central (optimal) values of $d_3^\text{est}$
in the table and figure carry important information
on the respective assumptions.
We should note that the errors of $d_3^\text{est}$
are only systematic 
and have no statistical nature. 
Hence, by carefully contemplating on the 
origins of these systematic errors, we
can extract the sizes and signs of the systematic effects.
The agreement with respect to the systematic errors is 
a necessary condition for the validity of our analysis given
below.

In the cases $n_l=3,4,5$, corresponding to
$c\bar{c}$, $b\bar{b}$, $t\bar{t}$ quarkonium
states, respectively, we see a good agreement of 
the estimates by 
\cite{Sumino:2013qqa} with the exact values, whereas the estimates by
\cite{Ayala:2014yxa} are slightly larger.
On the other hand, for smaller $n_l=0,1$, which correspond to 
hypothetical heavy quarkonium systems,
the agreement between the estimates by
\cite{Ayala:2014yxa} and the exact results is fairly good, whereas
the estimate by \cite{Sumino:2013qqa} for $n_l=0$ is slightly smaller than
the exact value.
From
these observations we derive the following interpretation:
\begin{itemize}
\item
For $n_l=3,4,5$, 
(i)
US corrections in $V_{\rm QCD}(r)$ are small, and
(ii) there is a stronger cancellation of IR contributions
than what has been predicted by 
renormalon dominance hypothesis.
\item
For $n_l= 0$,
(iii)
renormalon dominance holds more accurately, 
and 
(iv) non-negligible contributions from
US corrections exist.
\end{itemize}
We explain the details in the following.

The renormalon dominance hypothesis assumes that 
the expansion coefficient of the perturbative series
is dominated by a 
factorial ($\sim n!$) growth \cite{Beneke:1994rs},
\bea
d_n 
 \sim {\rm const.}\times  
\left(2{\beta_0}\right)^n \frac{\Gamma(n+\nu+1)}{\Gamma(\nu+1)}
~~~~~{\rm for}~~~~~n\gg 1,
\label{asymptform}
\eea
which stems from the singularity at $u=1/2$ in the
Borel transform of the perturbative series.
[\,$\nu=\beta_1/(2\beta_0^2)$, and
$\beta_i$ denotes the $(i+1)$-loop coefficient of the
beta function of $\alpha_s$.\,]
Contributions from the analytic part at $u=1/2$ are neglected.
The comparison between $d_3^\text{exact}$ and the
central values of $d_3^\text{est}[4]$ shows
that the renormalon dominance
hypothesis works better for smaller $n_l$.
This suggests that
the above factorial growth overwhelms contributions
from the analytic part as
$\beta_0(>0)$ becomes larger for smaller $n_l$.

\begin{figure}[h]
\begin{center}
\includegraphics[width=11cm]{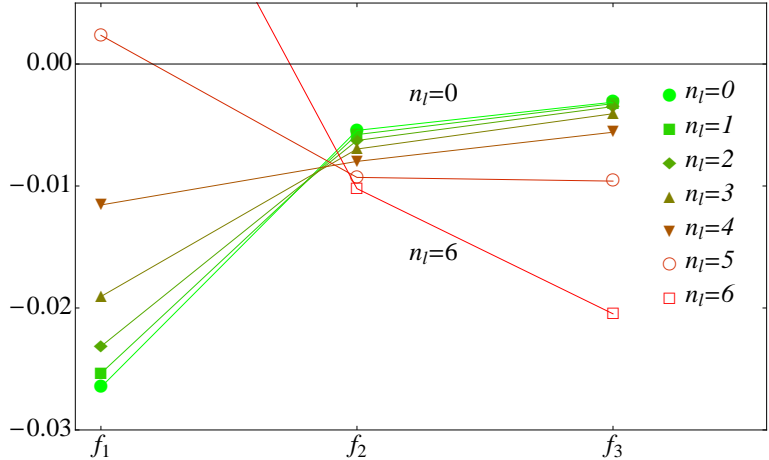}
\end{center}
\vspace*{-5mm}
\caption{\small
Each term of eq.~(\ref{1/Nexpansion}) for
$n=3$ and different $n_l$'s:
$f_1=\frac{\nu}{n+\nu}\,\tilde{c}_1 $,
$f_2=\frac{\nu(\nu-1)}{(n+\nu)(n+\nu-1)}\,\tilde{c}_2 $,
$f_3=\frac{\nu(\nu-1)(\nu-2)}{(n+\nu)(n+\nu-1)(n+\nu-2)}\,\tilde{c}_3 $.
The first term ($f_0=1$) is omitted, since it is
by far greater.
\label{tildeck}}
\end{figure}

Another source of $n_l$ dependence of the
renormalon dominance resides 
in the series \cite{Beneke:1994rs,Beneke:1998ui}
\bea
&&
F_n=
1
+
\frac{\nu}{n+\nu}\,\tilde{c}_1 
+
\frac{\nu(\nu-1)}{(n+\nu)(n+\nu-1)}\,\tilde{c}_2 
+
\frac{\nu(\nu-1)(\nu-2)}{(n+\nu)(n+\nu-1)(n+\nu-2)}\,\tilde{c}_3 
\nonumber\\ &&
~~~~~~~~
+ {\cal O}\left(\frac{1}{n^4}\right)
\label{1/Nexpansion}
\eea
in eq.~(33) of \cite{Ayala:2014yxa}.
The factor $F_n$ multiplies the right-hand side of
eq.~(\ref{asymptform}), giving $1/n$ suppressed
corrections, so that it shows how the expansion coefficient
approaches the asymptotic form at large
orders ($n\gg 1$).\footnote{
Contributions from the analytic part at $u=1/2$ are
not included in the series $F_n$, as they are suppressed
 exponentially.
}
Fig.~\ref{tildeck} plots the series (\ref{1/Nexpansion}) in
our case $n=3$ for different $n_l$'s.
They exhibit the tendency
that $1/n$ suppressed contributions become more important for larger $n_l$,
although the first term (=1) is by far dominating.
Both of these $n_l$ dependences in
analytic and $1/n$ suppressed contributions
 have been taken into account
in the error estimates of \cite{Ayala:2014yxa}.
The former error enters as scale dependences in the analysis
of \cite{Ayala:2014yxa} and is the main source of
errors.

One may wonder if the UV renormalon at $u=-1$ contained in the pole mass 
gives a significant contribution to the perturbative series of the pole mass.
Based on an analysis in the large-$\beta_0$ approximation,
we estimate that the contribution of the $u=-1$ renormalon to $d_3$ 
is fairly small compared to the errors of $d_3^\text{est}$\cite{Ayala:2014yxa}
listed in Tab.~\ref{results}.
This is consistent, since the analytic part
at $u=1/2$ contributes dominantly to these errors, and
the $u=-1$ renormalon  belongs to the analytic part.
The analysis also suggests that the $u=-1$ renormalon contribution
is not a dominant component 
of the analytic part.
Another important feature is that, since the UV renormalon is Borel
summable and gives a well-defined contribution, as long as we
obtain a converging series of a physical observable
(such as the heavy quarkonium energy level), 
the contribution of the $u=-1$ renormalon to the error estimate
becomes small (arbitrarily small unlike IR renormalons).  
Indeed contribution to the error is minor
at our present perturbative order.
We give details of the analysis of the $u=-1$ UV renormalon in
the Appendix.

Similarly there may be effects by the $u=1$ IR renormalon contained 
in the pole mass, whose properties are less known.
Known properties are as follows \cite{Neubert:1996zy}.
(a) It is induced by the non-relativistic kinetic energy
operator $\vec{D}^2/(2m)$;
(b) It is not forbidden by any symmetry, and 
parametrically it possibly induces an order $\LQ^2/m$ uncertainty;
(c) It does not appear in the large-$\beta_0$ approximation.
With this limited knowledge,
it is not easy to estimate contribution of the $u=1$ renormalon
in the estimate of $d_3$ in \cite{Ayala:2014yxa}.
In principle, this contribution is exponentially suppressed in the 
estimate of the $u=1/2$ renormalon in the pole mass and is encoded in 
the scale dependence in the error estimate of 
$d_3^\text{est}[4]$.

We turn to
the estimates of $d_3$ by \cite{Sumino:2013qqa}, which incorporate
the fact that cancellation of IR dynamics occurs beyond 
the renormalon dominance hypothesis.
It can be understood using the potential-NRQCD effective field 
theory \cite{Brambilla:1999xf}, 
in which interactions
of a heavy quarkonium and IR degrees of freedom are
systematically organized in multipole expansion in $r$.
The leading order interaction is given by the interaction
of an IR gluon with the total color charge of the heavy quarkonium,
which vanishes for a color-singlet system.
The corresponding
contribution to the binding energy is given by an
$r$-independent IR part of $2m_{\rm pole}+V_{\rm QCD}(r)$
\cite{Pineda:id}.
The cancellation between
$2m_{\rm pole}$ and $V_{\rm QCD}(r)$
is not restricted to the
renormalon part, and the analytic part
at $u=1/2$ contains such contributions.

In this general framework,
the lowest order non-canceled IR contribution to
the energy is given by
a double insertion of the dipole interaction between
the color-electric field and heavy quarkonium,
expressed in terms of
a non-local gluon condensate of the form
$\sim \langle\, \vec{r}\!\cdot\! \vec{E}^a\,
\vec{r}\!\cdot\! \vec{E}^a \rangle$.
It is dominated by
contributions from the US energy scale, and
the perturbative evaluation of this condensate
at ${\cal O}(\alpha_s^4\log\alpha_s)$
and ${\cal O}(\alpha_s^4)$ has been incorporated
in $V_{\rm QCD}(r)$ in the estimate of $d_3$.
In principle,
the $u=1$ renormalon in the pole mass
(if it exists) 
can affect 
$2m_{\rm pole}+V_{\rm QCD}(r)$.
However, the large mass limit $m\to \infty$
is taken in the analysis of
\cite{Sumino:2013qqa}, so that the $u=1$ renormalon
(order $\LQ^2/m$) is suppressed compared to the
$u=3/2$ renormalon (order $r^2 \LQ^3$).
Hence, the estimate of $d_3^\text{est}[3]$ should not be affected by the 
$u=1$ renormalon.
Furthermore, in estimating $d_3$ the effects of taking the
large mass limit are small for $n_l=3,4,5$, compared to the
real $c$, $b$, $t$-quark mass cases, hence, our discussion
is expected to be valid for these real heavy quarkonium
systems.
In the cases $n_l\le 2$ and $n_l\ge 6$, perturbative analysis makes sense
only in a hypothetical static limit $m\to \infty$, and our
discussion is confined to this limit.

In perturbative QCD, instability against
scale variation in IR region
is manifest for all the physical
observables, reflecting the blow-up of the
running coupling constant at IR.
For a ``good'' observable, generally
scale dependence decreases as the order of
perturbative expansion is raised.
Empirically this happens not only in the ultra-violet
(UV) direction but also stability extends to
IR region as the perturbation order is increased.
In the case of the heavy quarkonium energy,
the leading source of IR instability is the
non-local gluon condensate dominated by
US corrections.
The (optimal) values of the
estimates of $d_3$ in the first line of
Tab.~\ref{results} are chosen to optimize
the stability of the perturbative prediction for the energy
in the IR region at NNNLO.
A very good coincidence of these values
with the exact results for
$n_l=3,4$ suggests that the US corrections are
small for these systems.
Here, we may set the criterion for ``large'' or ``small'' by whether 
the corrections
deteriorate stability of the
perturbative prediction or not.

\begin{figure}[h]
\begin{center}
\includegraphics[width=10cm]{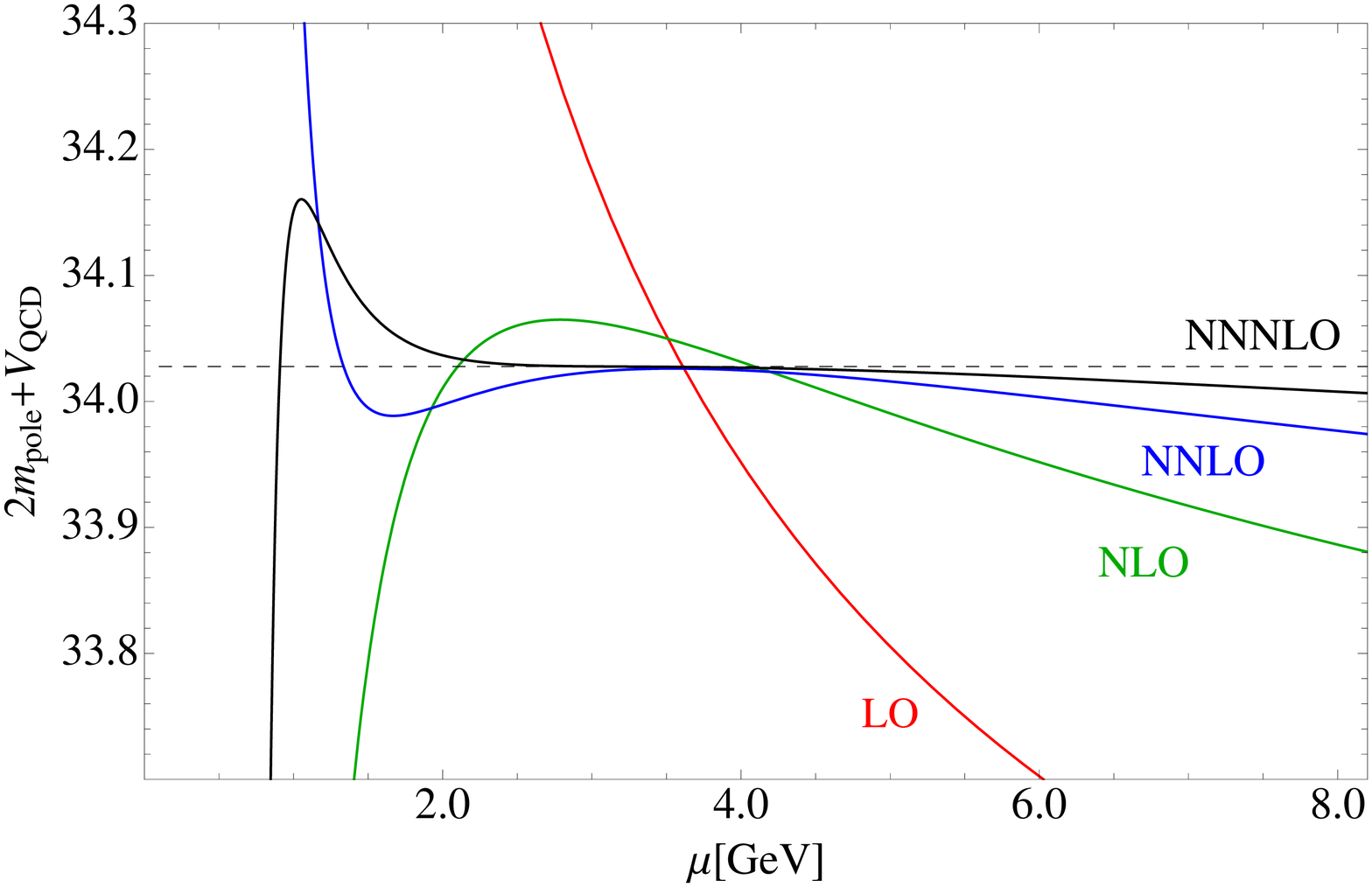}
\\
\vspace*{-3mm}
\end{center}
\caption{\small
Scale dependences of 
$2m_{\rm pole}+V_{\rm QCD}(r)$ at different orders
of perturbative expansion in the case $n_l=0$.
Input parameters are 
$\alpha_s(3\,\text{GeV})=0.2$ [$\alpha_s(\mu)$ blows up at
$\mu\approx 0.62$~GeV], $\overline{m}=16$~GeV
(a large value is chosen to suppress sub-leading renormalons
in $m_\text{pole}$), and
$r=0.5~\text{GeV}^{-1}$.
The exact value of $d_3$ is used.
A horizontal line is shown as a guide.
\label{Instability}}
\end{figure}

As shown in \cite{Sumino:2013qqa}, 
perturbative stability of $2m_{\rm pole}+V_{\rm QCD}(r)$
is sensitive to the precise value of $d_3$, and
this sensitivity turns out to be asymmetric with respect to
the sign of a variation of $d_3$.\footnote{
Qualitatively the same feature is observed for the
heavy quarkonium energy levels \cite{Kiyo:2013aea}.
}
If $d_3$ is larger than a certain critical value, stability
of the prediction is lost very quickly.
This leads to a fairly sharp upper bound on the estimate of
$d_3$ for each $n_l$.
By contrast, stability of the prediction is degraded only gradually
if $d_3$ is lowered from its optimal value.
In this regard,
a marked result is that in the case $n_l=0$
the exact value
of $d_3$ is on the verge of or slightly above
the upper bound of $d_3$ required by stability of the energy.
Since US corrections are expected to be the source of IR
instability of the energy,
we infer that the US corrections are sizable in this case.
Oppositely, in the case $n_l=5$, the exact value of $d_3$ lies slightly
below that required by optimal stability of the
energy.
Hence, in this case US corrections do not 
deteriorate perturbative stability in any essential way,
and US corrections may well be regarded as ``small.''
Such a dependence of IR stability
on $n_l$ may result from the fact that
the running coupling constant blows up most rapidly for
$n_l=0$, while the running becomes milder as $n_l$ increases.
(Note that we consider $n_l$ massless quarks.)
If an IR catastrophe of perturbative stability should ever occur,
it would be expected to appear first in
the most rapidly running case.
To demonstrate explicitly the level of instability in the case $n_l=0$, we show
a plot according to the analysis of \cite{Sumino:2013qqa}.
Fig.~\ref{Instability} shows the scale dependences of
$2m_{\rm pole}+V_{\rm QCD}(r)$ at a relatively large $r$, where perturbative
stability up to NNLO is close to marginal.
The NNNLO line is flatter than the NNLO line in the 
large $\mu$ region, however,
it grows in the small $\mu$
region and starts to show a sign of instability.
See \cite{Sumino:2013qqa} for more details of the analysis method.

There is a difficulty in quantifying the size of
US corrections more directly.
By definition, the US corrections are dependent on the
factorization scale $\mu_f$, which should satisfy the condition
\cite{Brambilla:1999xf}
\bea
\LQ, \, \frac{C_A\alpha_s}{2\,r}\ll \mu_f \ll \frac{1}{r}  ,
\eea
where $C_A=N_C=3$ is the Casimir operator for the adjoint representation.
Given the different $n_l$ dependences of 
$d_3^\text{est}\cite{Sumino:2013qqa}$ and 
$d_3^\text{exact}\cite{Marquard:2015qpa}$,
we confirm that a simple logarithmic dependence of the US corrections
on $\mu_f$, proportional to
$\alpha_s^5 \log (\mu_f r)$, cannot explain the
difference, even if we
assume a reasonable $n_l$ dependence of $\mu_f$.
This is expected, since except in dimensional regularization,
which conceals power-like dependences on the scales,
we expect a much stronger dependence $\sim \mu_f^3 r^2$ of
the US corrections.
This dependence should eventually turn into a dependence
on the physical US scales, namely $\mu_f$ should
be replaced by $C_A\alpha_s/(2r)$ and $\LQ$,
where presumably the latter is more dominant at larger $r$.\footnote{
$\mu_f$ dependence is canceled in physical observables.
Hence, we are ultimately interested in the dependence of physical 
observables on the physical US scale.
At lower orders of perturbative series, only the
scale $C_A\alpha_s/(2r)$ is visible.
As the order is raised, perturbative expansion becomes more
sensitive to the $\LQ$ scale.
The leading dependence of $2m_{\rm pole}+V_{\rm QCD}(r)$
on $\LQ$ should appear as $\LQ^3r^2$.
}
This requires (at least)
an analysis analogous to that of \cite{Ayala:2014yxa}
incorporating the $u=3/2$ singularity in addition.
Furthermore, we would need to separate UV and IR contributions
in perturbative expansion systematically, to be able to
accurately extract the US contributions \cite{Sumino:2005cq,Sumino:2014qpa}.
Such a detailed analysis is beyond the scope of this paper.
\clfn

Thus, for the case $n_l=0$, 
we are (for the time being) 
content with the observation that everything
is consistent.
The renormalon
dominance hypothesis can accurately estimate $d_3$
by the method of \cite{Ayala:2014yxa}.
As mentioned, it is plausible that the
renormalon dominance works most accurately
in this case.\footnote{
This feature appears to be slightly
reinforced for $n_l=0$ by a cancellation of
the contribution from the $u=-1$ UV renormalon and
other contributions from the analytic part at $u=1/2$;
compare Tabs.~\ref{results} and \ref{UVrenInd3}.
}
On the other hand, 
stability of  $2m_{\rm pole}+V_{\rm QCD}(r)$ at IR
can in principle be jeopardized by US corrections
and, if at all, this is expected to happen for smaller $n_l$.

In contrast, for $n_l=3,4,5$, the analytic part at $u=1/2$ has a larger
relative significance, and the central values of the estimates
by \cite{Ayala:2014yxa} depart from the exact values;
see Tab.~\ref{results} and Fig.~\ref{ErrorBars}.
We can circumvent
this problem in the method of \cite{Sumino:2013qqa},
since cancellation of IR dynamics takes place in
the analytic part as well, and the contribution of US
corrections is expected to be milder than the $n_l=0$ case.
Thus, we are led
to the interpretation as presented in the beginning
of this discussion.


On the basis of our understanding up to this point,
we reexamine the prediction of the energy level of the
(would-be) toponium $1S$ state, using the NNNLO formula for the $1S$ energy
level \cite{Beneke:2005hg,Penin:2002zv,Kiyo:2014uca}.
We compare with the analysis \cite{Kiyo:2002rr}, which examined
the $1S$ energy level calculated in terms of 
$\overline{m}_t\equiv m_t^{\overline{\rm MS}}(m_t^{\overline{\rm MS}})$
and in the $\varepsilon$ expansion
\cite{Hoang:1998ng}.
The large-$\beta_0$ approximation (a crude approximation based on
renormalon dominance) was used for estimates of $a_3$ and $d_3$.\footnote{
More accurately, a Pad\'e estimate of
$a_3$ was used and the prediction of the $1S$ energy level 
was shown to be quite close to that of the large-$\beta_0$
approximation.
Since the difference is minor and
irrelevant in our context, we refer only to the 
large-$\beta_0$ approximation.
}
We replace them by the
exact values.
The essence of the analysis 
\cite{Kiyo:2002rr} is to use the renormalon dominance hypothesis
for estimating a perturbative error in the
top quark mass determination from the energy level of the
toponium $1S$ state.
As a result, about 40~MeV for an expected accuracy was predicted
for determination of the top quark $\overline{\rm MS}$ mass.

\begin{figure}[h]
\begin{center}
\includegraphics[width=10cm]{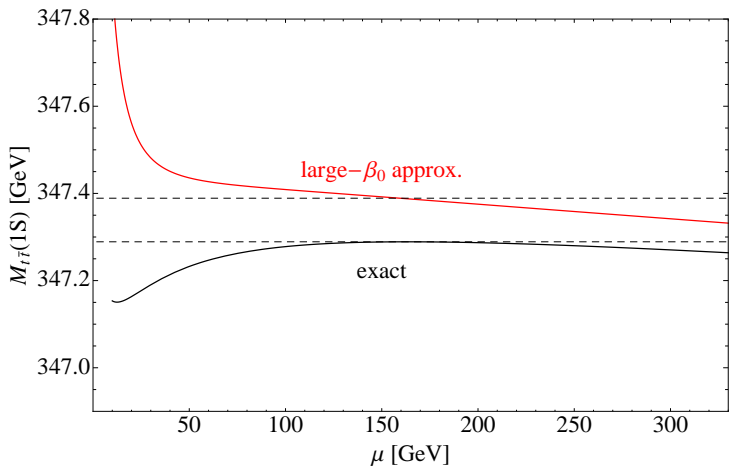}
\\
\vspace*{-3mm}
\end{center}
\caption{\small
Comparison of scale dependence of the toponium $1S$ energy
at NNNLO from the previous analysis
(the large-$\beta_0$ approximation \cite{Kiyo:2002rr})
and that using the exact values of $a_3$ and $d_3$.
The input values are $\overline{m}_t=165$~GeV,
$\alpha_s(M_Z)=0.1185$, and $n_l=5$.
Horizontal dashed lines are shown as a guide.
\label{tt1S-exact.vs.largebeta0}}
\end{figure}

All the qualitative argument of \cite{Kiyo:2002rr}
based on renormalon dominance hypothesis should be valid, since,
as we have verified,
the renormalon dominance is qualitatively a good approximation.
Nevertheless,
according to our above understanding, the accuracy of the prediction
is expected to improve, since the cancellation of IR dynamics
occurs at a deeper level than that of the large-$\beta_0$
approximation.
In the $t\bar{t}$ system, the leading non-canceled IR contribution
from US corrections is expected to be ``small'' if our
understanding is consistent.

Fig.~\ref{tt1S-exact.vs.largebeta0} compares
the scale dependence of the toponium $1S$ energy by the
previous analysis \cite{Kiyo:2002rr}
and that using the exact values of $a_3$ and $d_3$.
A marked difference is that the former prediction is
much more unstable in the IR region than the latter.
This is consistent with our expectation.
There also appears a flat region (minimal-sensitivity scale 
\cite{Stevenson:1981vj})
in the new prediction, which is absent in the former
prediction.

\begin{figure}
\begin{center}
\includegraphics[width=12cm]{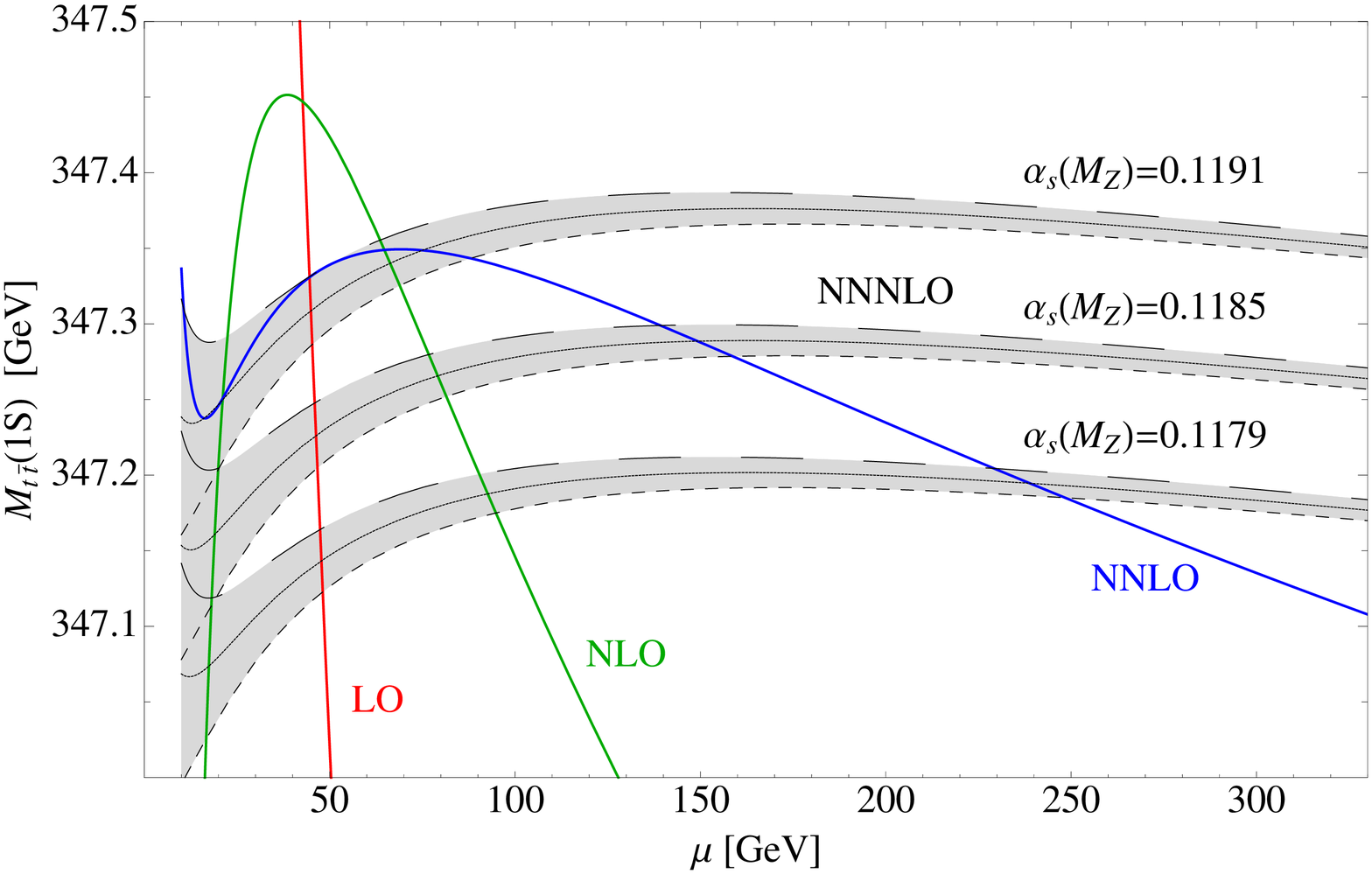}
\\
\caption{\small
Scale dependence of the toponium $1S$ energy level.
The input $\overline{\rm MS}$ mass is taken as $\overline{m}_t=165$~GeV.
Each band for the NNNLO prediction 
corresponds to variation of $d_3^\text{exact}$
inside its error ($\pm 21.5$),
where the upper (lower) line in each band corresponds to the upper
(lower) value of $d_3^\text{exact}$.
The different bands correspond to
different input values of $\alpha_s(M_Z)$.
Predictions at lower orders are for $\alpha_s(M_Z)=0.1185$.
\label{tt1S}}
\vspace*{10mm}
\includegraphics[width=12cm]{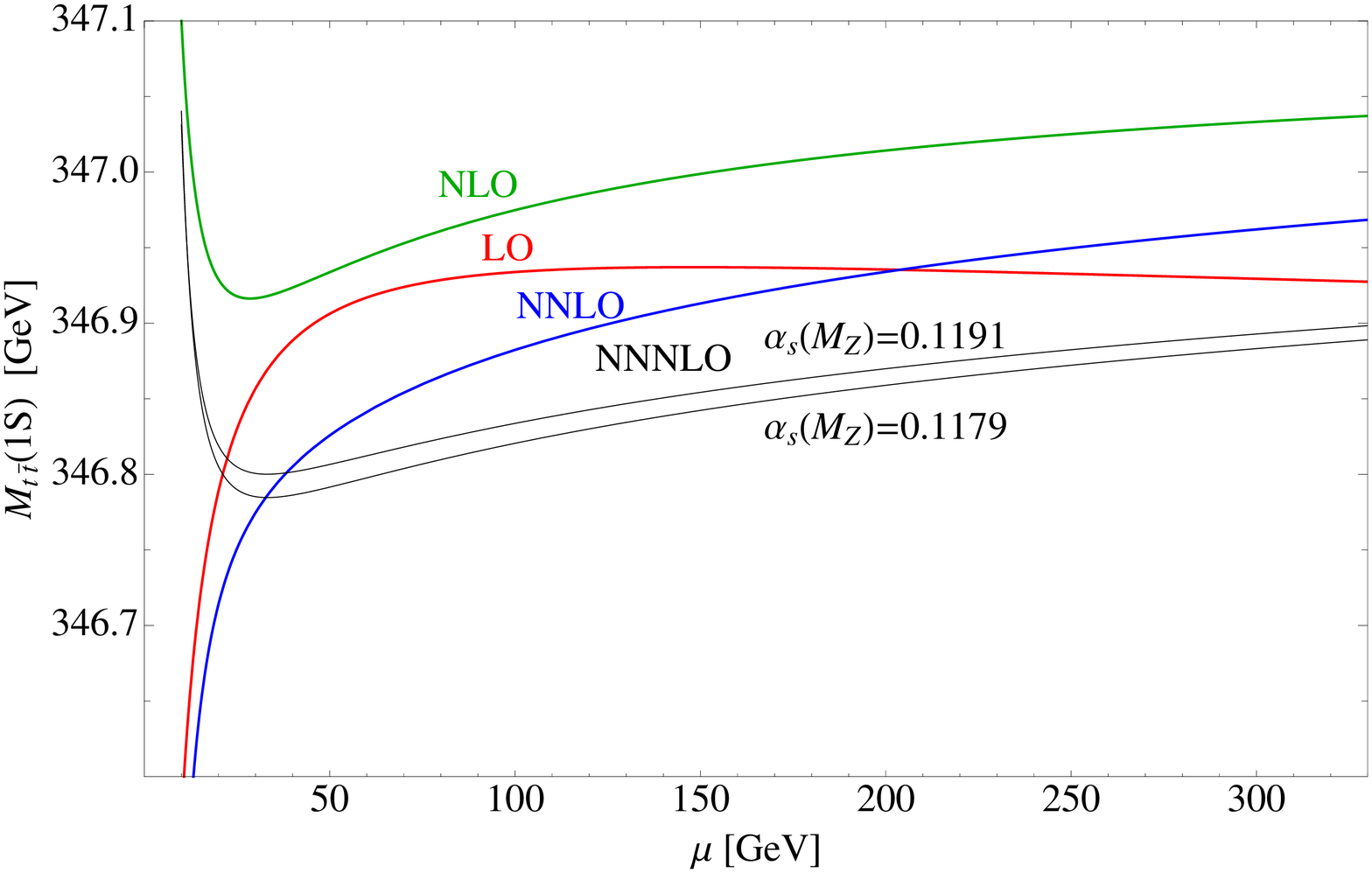}
\caption{\small
\label{figPSmass}
Scale dependence of the $1S$ energy level in the PS--mass scheme
at different orders.  
The input PS mass is taken as 
$m_{\rm PS}(\mu_{f,{\rm PS}}=20{\rm GeV})=173\, {\rm GeV}$. 
Two lines for the NNNLO result correspond
to $\alpha_s(M_Z)=0.1179$ and $0.1191$.
At lower orders 
$\alpha_s(M_Z)=0.1185$ is used.
}
\end{center}
\end{figure}

We estimate the error of the new prediction.
It is natural to use the scale dependence around
the minimal-sensitivity scale 
($\approx 160$~GeV).\footnote{
From the general argument based on the renormalon dominance
hypothesis,
the minimal-sensitivity scale is expected to increase
as the perturbation order is raised \cite{Sumino:2005cq}; see
Fig.~\ref{tt1S}.
}
Following the standard prescription we vary the scale by
factors 1/2 and 2.
When the scale $\mu$ is varied between
80 and 320~GeV,  
the $1S$ energy varies by
about 20~MeV below and above the minimal-sensitivity
scale, respectively.
Therefore, the sum of the absolute variations 
of the $1S$ energy level
is about 40~MeV.\footnote{
This is a factor 2 more conservative estimate than taking
the maximal variation of the $1S$ energy level in this
range.
}
The corresponding
variation of the top quark $\overline{\rm MS}$
mass is almost one half of it, leading to about 20~MeV, which
we take as an error estimate.
Another error estimate may be obtained from
the difference between the NNLO
prediction at the minimal sensitivity scale (at NNLO)
and that at NNNLO, namely the difference
between the values of $M_{t\bar{t}}(1S)$ at the 
local maxima at NNLO and NNNLO in Fig.~\ref{tt1S}.
This gives 30~MeV as an uncertainty for the top quark mass.
For reference, we show the series expansion in $\varepsilon$ at the
minimal sensitivity scale at NNNLO:
\bea
M_{t\bar{t}}(1S)=2\times
(165 + 7.20 + 1.22 + 0.216 + 0.0077 )~\text{GeV}
~~~\mbox{for~~~$\mu=162$~GeV},
\eea
which shows a healthy convergence behavior
[$\overline{m}_t=165$~GeV and $\alpha_s(M_Z)=0.1185$].
Thus, we estimate an error in the top quark
$\overline{\rm MS}$ mass determination from
$M_{t\bar{t}}(1S)$ to be
20--30~MeV.

We note that the naive error estimate of
order $\LQ^3/(\alpha_s m)^2$ by the uncanceled renormalon
at $u=3/2$ in \cite{Kiyo:2002rr} is order 3--10~MeV, which is still 
somewhat smaller than the
current error estimate.
This means that the current perturbative order would not be high enough
to be limited by this renormalon uncertainty.
Contribution of the renormalon at $u=-1$ in the pole mass
is estimated to be a few MeV or less (see the Appendix), while
contribution of the $u=+1$ renormalon 
is estimated naively to be 
order $\LQ^2/m \sim 0.5$--$1.5$~MeV 
(corresponding to $\LQ \sim 300$--$500$~MeV).

In Ref.~\cite{Kiyo:2002rr} the range of the scale variation was 
taken differently 
from the above range, since no minimal-sensitivity
scale for the $1S$ energy exists for that prediction and
a different criterion was used.
We may check consistency.
If we vary the scale in the above range  for the previous prediction,
we obtain
the same error estimate for the top quark $\overline{\rm MS}$
mass as in \cite{Kiyo:2002rr}  (about 40~MeV).

Thus, we obtained a better possible accuracy of the
top quark mass determination at a future linear collider
over the previous estimate, which relied only
on the renormalon dominance
hypothesis before the full computations of $a_3$ and $d_3$.
We consider that it is not a sheer 
numerical accident but with a reasoning that we
obtain a smaller error estimate.
Namely, from the
general property of QCD a stronger IR cancellation 
than what is predicted by
the renormalon dominance hypothesis follows.
This interpretation is supported by a detailed 
comparison between the
estimates of $d_3$ for $n_l=3,4,5$ from stability of
$2m_{\rm pole}+V_{\rm QCD}(r)$ and
the estimates by the renormalon dominance, and also by an
overall consistent picture drawn in the first part
of this paper.

To clarify the current status, we show in Fig.~\ref{tt1S} dependences of
the $1S$ energy level 
on the current uncertainty of the exact value of $d_3$ and
on the input value of $\alpha_s(M_Z)=0.1185 \pm 0.0006$ \cite{Agashe:2014kda}.
The former induces about 10~MeV variation (5~MeV for the top quark
mass) at the minimal-sensitivity scale, while
the latter induces about 90~MeV (45~MeV for the top mass) variation.
Hence, a precise determination of $\alpha_s(M_Z)$, of the
order of $\pm 0.0001$ accuracy,
is prerequisite
to achieve 20--30~MeV accuracy of the top quark
mass determination.
Prediction of $d_3$ with higher precision is also favorable.

For comparison,
we perform a similar analysis using  the potential subtracted (PS) mass 
\cite{Beneke:1998rk} as the input parameter.
(The definition of the NNNLO PS mass is given
in \cite{Beneke:2005hg}.)
Fig.~\ref{figPSmass} shows the scale dependence of the toponium $1S$ 
energy level, where 
we use the PS mass $m_{\rm PS}(\mu_{f,{\rm PS}}=20\,{\rm GeV})=173\, {\rm GeV}$.
To compare with the $\overline{\rm MS}$ mass,
we vary the scale from $80\,{\rm GeV}$ to
 $320\,{\rm GeV}$ and find the variation of the
$1S$ energy level of about $75\,{\rm MeV}$.
(For $50\,\text{GeV}\leq \mu \leq 350 \, {\rm GeV}$, 
the variation is about 
$100\, {\rm MeV} $,
which is consistent with \cite{Beneke:2015zqa}.)
The uncertainty of   $\alpha_s(M_Z)$ 
causes $\pm 8\,{\rm MeV}$ shift 
of the NNNLO energy level.\footnote{
The dependence of the PS mass on $\alpha_S(M_Z)$ starts from the
order $\alpha_s^2$, which is the reason for a smaller dependence 
compared to the $\overline{\rm MS}$ mass.
}
Thus, use of the PS mass leads to
a larger scale variation of the perturbative prediction for
the $1S$ energy level
than the $\overline{\rm MS}$ mass.
We observe qualitatively different scale dependences
between the two schemes by comparing Figs.~\ref{tt1S}
and \ref{figPSmass}, where this tendency is apparent
not only at NNNLO but also at lower orders.
Furthermore, we confirm a similar tendency in the
scale dependences for other $n_l$'s, where the 
values of $d_3$ vary considerably.  
We also note that the conversion formula between the 
PS and $\overline{\rm MS}$ masses
induces a scale uncertainty of order 30\,MeV for 
$80\,\text{GeV} < \mu < 320\,\text{GeV}$
provided that  $m_{\rm PS}=173$\,GeV is an input value.

Intuitively the difference between using the 
$\overline{\rm MS}$ and PS masses may be understood 
as follows.
In the $\overline{\rm MS}$--mass scheme, the energy of the
toponium bound state consists of 
(i) the $\overline{\rm MS}$ masses of $t$ and 
$\bar{t}$, (ii) contributions to the self-energies of $t$ and 
$\bar{t}$ not renormalized 
into the $\overline{\rm MS}$ mass (typically from gluons whose wavelengths 
are larger than the Compton wavelength of $t$, 
$\lambda_g \simgt 1/m_t$), and 
(iii) the potential energy between $t$ and 
$\bar{t}$.
IR contributions between (ii) and (iii) (typically from 
$\lambda_g$ larger than the 
bound-state size) get canceled, where the domain of IR cancellation 
is determined dynamically by the wave function of the bound 
state \cite{Brambilla:2001fw,Recksiegel:2002za,Sumino:2014qpa}.

The composition of the energy of the bound state
in the PS--mass scheme is similar, except that the 
renormalized mass (i) is replaced by the PS mass, which renormalizes the
top quark self-energy from $\lambda_g \simlt 1/\mu_{f,{\rm PS}}$.  
In the computation of the self-energy, a sharp cut-off is
introduced  in momentum space at the factorization scale $\mu_{f,{\rm PS}}$,
which is chosen to be of the order of the Bohr 
scale $\sim\alpha_s m_t$.  
The cut-off induces a power dependence of the
PS mass on $\mu_{f,{\rm PS}}$. 
Since the $1/\mu_{f,{\rm PS}}$ is close to the bound-state size, the IR 
cancellation can become incomplete by artificial cut-off effects if $\mu_{f,{\rm PS}}$ is 
too low.  
Such effects tend to be enhanced, due to the increase of the coupling 
constant at IR and the power dependence on $\mu_{f,{\rm PS}}$.

\begin{figure}
\begin{center}
\includegraphics[width=7.7cm]{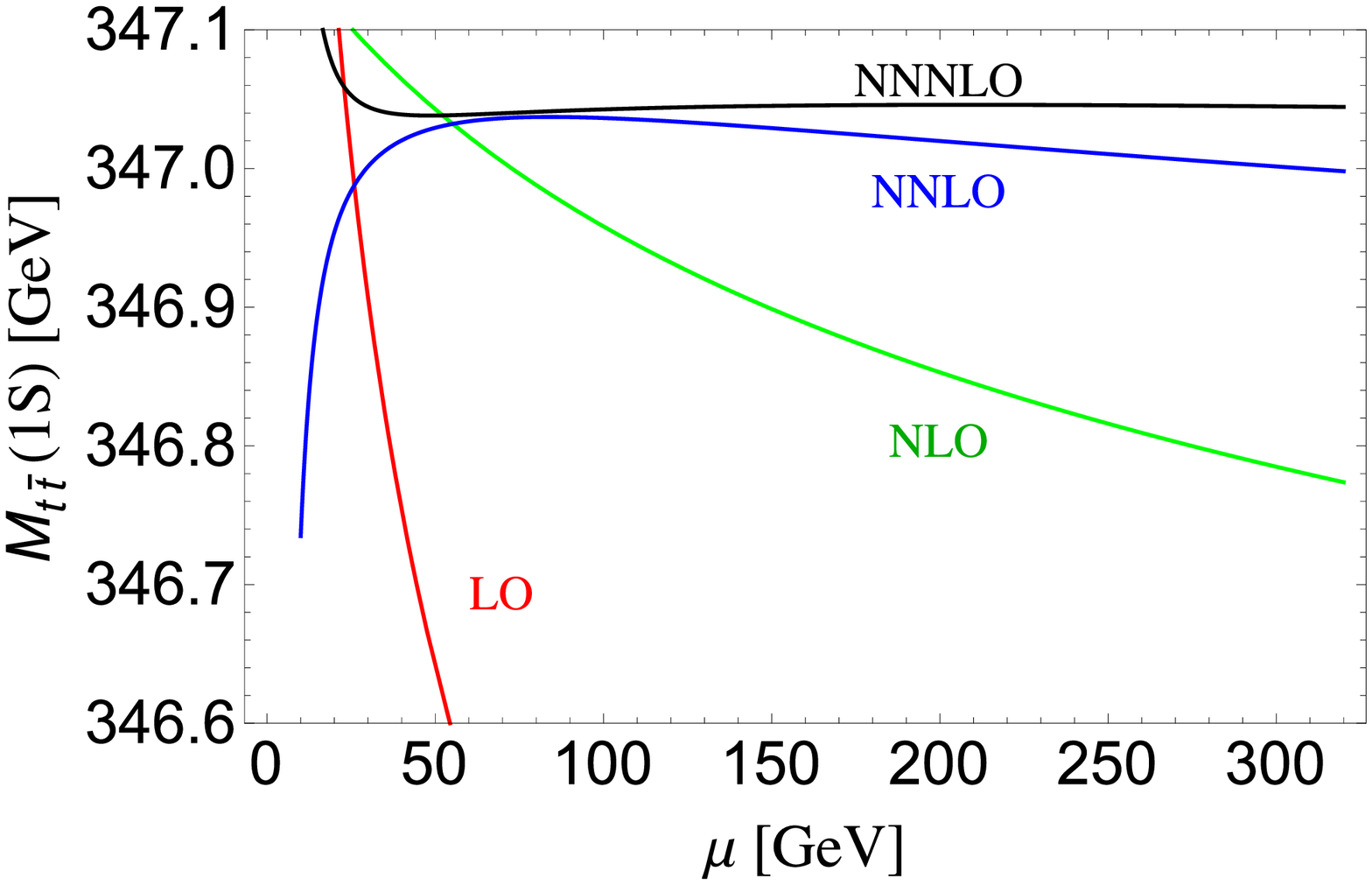}
\includegraphics[width=7.7cm]{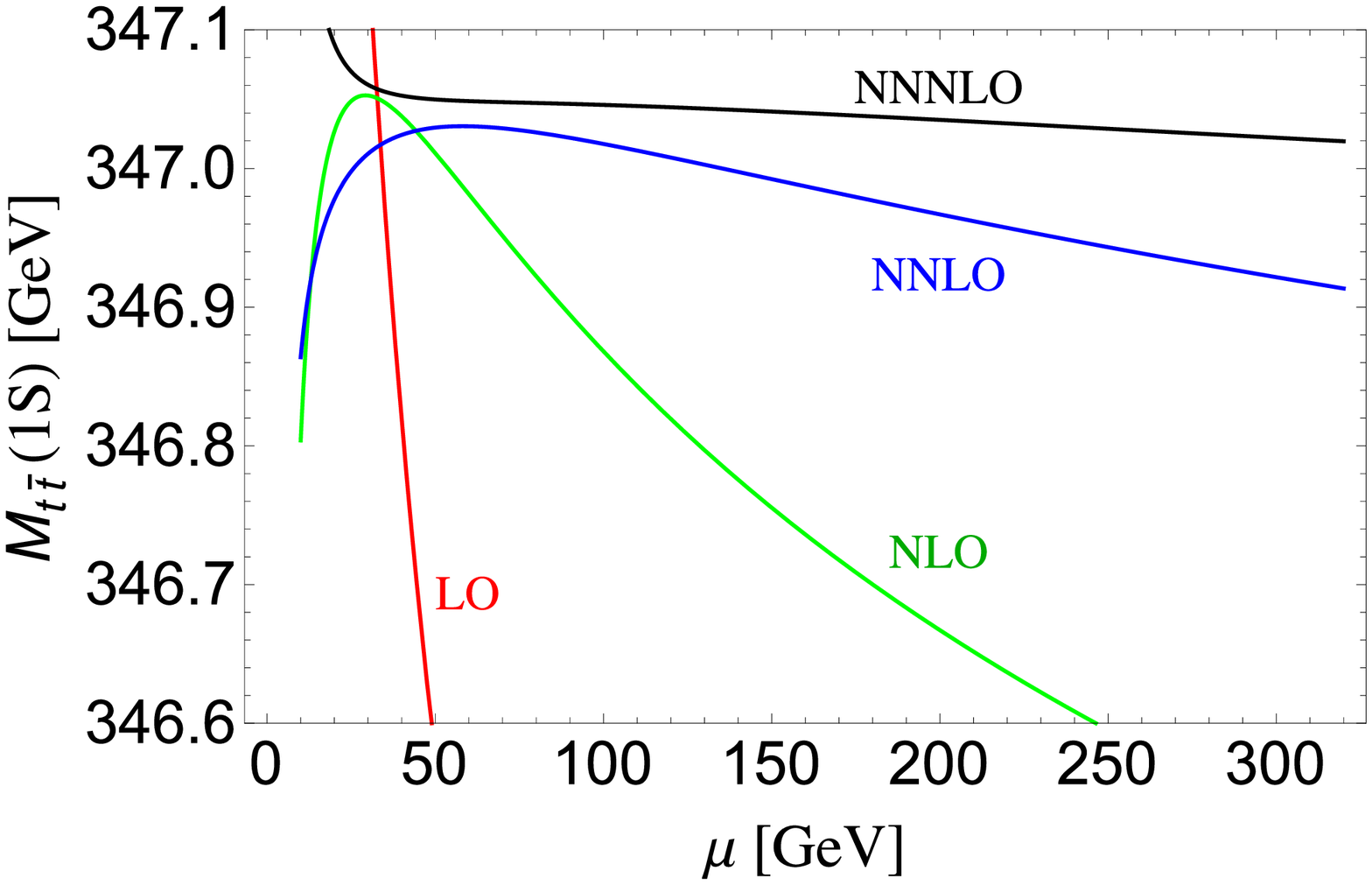}
\\
(a)~~~~\hspace*{67mm}(b)
\caption{\small
\label{figPSmass-muf=50+80}
Scale dependence of the $1S$ energy level in the PS--mass scheme
at different orders in the cases
(a) $\mu_{f,{\rm PS}}=50~{\rm GeV}$, $m_{\rm PS}=171.2$~GeV, and 
(b) $\mu_{f,{\rm PS}}=80~{\rm GeV}$, $m_{\rm PS}=169.5$~GeV. 
The values of the PS mass are chosen such that
the lines fit in the same range as in Fig.~\ref{figPSmass}.
}
\end{center}
\end{figure}

We may check consistency of this picture, by computing the energy level
in the case that $\mu_{f,{\rm PS}}$ 
is taken to be larger than the Bohr scale.\footnote{  
In principle this is at odds with 
the standard counting of $\varepsilon$ in the PS--mass scheme.  
Furthermore, the approximation
of subtracting the IR part of the pole mass
by an integral of $-V_{\rm QCD}(q)/2$ becomes worse 
as $\mu_{f,{\rm PS}}$ approaches $m_t$.
Hence, we take the cut-off in the range
$\alpha_s m_t < \mu_{f,{\rm PS}}< m_t$.
}  
In this case, the behavior of the predictions in the PS--mass scheme 
is expected to approach
qualitatively that of the $\overline{\rm MS}$--mass scheme,
as only shorter-wavelength contributions are renormalized in the PS mass
and IR cancellation becomes more complete
(artifact of cut-off diminishes).
We show in Figs.~\ref{figPSmass-muf=50+80}(a)(b) the energy level for
$\mu_{f,{\rm PS}}=50$ and 80~GeV, to be compared with Figs.~\ref{tt1S},
\ref{figPSmass},
and confirm this tendency. 
(We confirm qualitatively similar behavior for the bottomonium 
energy level as well.)

Let us discuss other sources of errors.
Besides what we have analyzed here,
there are many sources of uncertainties,
both of theoretical and experimental origins, in the actual
top quark mass determination at ILC.
Theoretically, these include
effects of 
mixed electroweak and QCD corrections
(finite width corrections, non-resonant diagrams,
non-factorizable corrections, etc.),
uncertainties in the normalization  and shape of the
threshold cross section,
contributions from higher-spin quarkonium states,
method for smooth matching to the high-energy cross section,
and so forth.
In addition effects of the initial-state radiation and 
beam energy spread need to be
taken into account in a realistic experimental situation
for the top quark threshold scan.  
(See \cite{Martinez:2002st, Horiguchi:2013wra} for recent simulation 
studies for the threshold scan at ILC.) 
Since feasibility of a high precision top quark
mass determination 
can be addressed only by  realistic simulation studies
incorporating all the above effects,
the accuracy we present here is what can be
achieved in principle, as a limitation from perturbative
QCD.
Nevertheless, such a precision is a unique possibility achievable
only at a future $e^+e^-$ collider and worth pursuing.

\section*{Note Added:}

After we completed our work, an analysis was reported
on the top quark mass determination using the NNNLO
$t\bar{t}$ cross section near threshold and
using the PS mass \cite{Beneke:2015zqa}.
Their estimate of about 50~MeV accuracy is larger than
the estimate presented in this paper (20--30~MeV), which is based only on
the uncertainty of the $1S$ energy level using the $\overline{\rm MS}$ 
mass.
Currently it remains an open question, in the case that
the cross section is computed thoroughly in terms of the 
$\overline{\rm MS}$ mass only, whether the latter estimate
is increased substantially due to
an uncertainty in the shape of the threshold 
cross section.

\section*{Acknowledgements}
The authors are grateful to M.~Beneke and M.~Steinhauser for 
useful comments
on our manuscript.
The authors also thank the editor and the referee for
bringing our attention to the $u=1$ renormalon and 
Ref.~\cite{Neubert:1996zy}.
The works of Y.K.\ and Y.S., respectively,
were supported in part by Grant-in-Aid for
scientific research Nos.\ 26400255 and 26400238 from
MEXT, Japan.
The work of G.M. is supported in part by Grant-in-Aid
 for JSPS Fellows (No.~26-10887).

\appendix
\section{UV renormalon}
\clfn 

\begin{table}[t]
\begin{center}
\begin{tabular}{c|ccccccc}
\hline 
$n_l$& 0 & 1 & 2 & 3 & 4 & 5 &6 \\
\hline 
$d_3^{[u=-1]}$ & 
31.4 & 26.1 & 21.3 & 17.2 & 13.7 & 10.6 & 8.1 \\
\hline
\end{tabular}
\caption{\small
Estimates by the large-$\beta_0$ approximation
for the 
contribution of the $u=-1$ UV renormalon to $d_3$.
}
\label{UVrenInd3}
\end{center}
\end{table}

In this appendix we estimate contributions 
to the pole--$\overline{\rm MS}$ mass relation 
from the UV renormalon 
at $u=-1$
using the large-$\beta_0$
approximation and estimate an uncertainty originating from this
renormalon.
Using the formula in \cite{Beneke:1994qe},
the contribution to $d_n$ [defined in eq.~(\ref{m-pole})] 
from the pole at $u=-1$  is given by 
\bea
d_n^{[u=-1]}= e^{-5/3}\,C_F (-1)^{n+1} \left(\frac{\beta_0}{4}
\right)^{n}  n! 
~~~~~(\mbox{large-$\beta_0$ approx.})
\, ,
\label{UVrenIndn}
\eea
where $C_F=4/3$ is the color factor.
In particular the contributions to $d_3$ are evaluated explicitly for
various $n_l$ in Tab.~\ref{UVrenInd3}.
Comparing these values with the corresponding errors of 
$d_3^\text{est}$\cite{Ayala:2014yxa} in Tab.~\ref{results},
we find that they are smaller than the errors
by factors 4--6 for $0\leq n_l \leq 4$
and by factors 10--20 for $n_l=5,6$.
This is consistent, since each error of
$d_3^\text{est}$\cite{Ayala:2014yxa} is dominated by the
contribution
from the analytic part at $u=1/2$ and
$d_3^{[u=-1]}$ belongs to the analytic part.
It suggests that the $u=-1$ UV renormalon is not a 
dominant component of the contribution from the analytic part
(for $n=3$).

In the rest of this appendix we estimate the contribution of the $u=-1$
UV renormalon to the quarkonium $1S$ energy level, taking
the toponium case ($n_l=5$) as an example.
(The case for the bottomonium is qualitatively similar.)

In the $1S$ energy level (at the leading-logarithms)
only the pole mass contains the
$u=-1$ UV renormalon.
In general a
UV renormalon induces
a factorial growth of perturbative series, 
as shown in eq.~(\ref{UVrenIndn})
(similarly to an IR renormalon), which breaks 
convergence of the perturbative series.
Nevertheless, since the corresponding singularity 
in the Borel plane ($u$--plane)
lies along the negative real axis, 
a definite value can be assigned to the contributions of 
a UV renormalon by Borel summation.
The perturbative series corresponding to 
a UV renormalon converges up to a certain order ($n<n_*$)
and diverges
beyond that order ($n>n_*$), which is
a typical feature of an asymptotic series.
In the case of the $u=-1$ UV renormalon 
(the UV renormalon nearest to the origin in the Borel plane),
the critical order $n_*$ is given by
\bea
n_* \approx \frac{4\pi}{\beta_0 \alpha_s(\overline{m}_t)} \approx 15 \, .
\eea
Therefore, the perturbative series is still converging in our NNNLO calculation.
The first several terms of the $u=-1$ contribution 
(in the large-$\beta_0$ approximation) read
\bea
&&
M_{t\bar{t}}(1S)^{[u=-1]}\equiv 2\,\overline{m}_t
\left[ 1+
\sum_{n=0}^\infty
d_n^{[u=-1]} \left( \frac{\alpha_s(\overline{m}_t)}{\pi} \right)^{n+1}
\right]
\label{defM1SUVren}
\\ &&
~~~~~~~~~~~~~~~~~
=
2\times[
165 - 1.44 + 0.096 -0.013 + 0.0025 - 0.00068 + 
\cdots
] ,
\label{seriesUVren}
\eea
for $\mu=\overline{m}_t=165$~GeV and $\alpha_s(\overline{m}_t)=0.109$.
According to a standard estimate with an asymptotic series,
the error of the prediction is of the order of the last known
term.
Hence, at NNNLO, we can estimate the error due to the
$u=-1$ renormalon to be of order 2.5~MeV for the top quark mass 
determination.

Alternatively we can estimate the error using 
the difference between the Borel summed value and the perturbative 
contribution up to NNNLO:
\bea
&&
\delta \overline{m}_t =
-\frac{4C_F\overline{m}_t}{e^\frac{5}{3}\beta_0} 
\int _0^\infty \! \! du
\left[ \frac{1}{1+u} -(1-u+u^2-u^3)\right] 
\exp \! \left[-\frac{4\pi u}{\beta_0 \alpha_S(\overline{m}_t)}  \right]
\nonumber\\
&&~~~~~
\approx -0.51~\text{MeV}.
\label{eq:UVrenormalon}
\eea
It is somewhat
smaller than the above estimate.
[An error estimate by the N$^4$LO term of eq.~(\ref{seriesUVren})
gives a better estimate.]

From the above examinations, one expects that the contribution of
the $u=-1$ renormalon is fairly modest and minor in the
error estimate in the determination of the top quark mass, which
is performed in the main body of this paper.
As long as the perturbative series is converging, the
error due to the $u=-1$ renormalon decreases.
This is in contrast to the $u=3/2$ renormalon, which induces a
limitation in achievable accuracy
of order $\Lambda_{\rm QCD}^3/ (\alpha_s m_t)^2$.
A crude estimate based on the large-$\beta_0$
approximation indicates that at NNNLO 
the error due to the $u=-1$ renormalon
is smaller than the error due to the $u=3/2$ renormalon.


\end{document}